\documentclass[runningheads]{llncs}
\usepackage[T1]{fontenc}
\usepackage{graphicx}
\usepackage{cite}
\usepackage{amsmath,amssymb,amsfonts}
\usepackage{textcomp}
\usepackage{xcolor}
\def\BibTeX{{\rm B\kern-.05em{\sc i\kern-.025em b}\kern-.08em
    T\kern-.1667em\lower.7ex\hbox{E}\kern-.125emX}}

\usepackage{tikz}
\usetikzlibrary{chains,positioning,arrows.meta,positioning,quotes,shapes}
\usepackage{float}
\usepackage{multicol}
\usepackage{array}
\usepackage{listings}
\usepackage{booktabs}
\usepackage{tabularx}
\usepackage{alltt}
\usepackage{adjustbox}
\usepackage{caption}
\usepackage{enumitem}
\usepackage{subcaption}
\usepackage{hyperref}

\usepackage{geometry}
\usepackage[T1]{fontenc}

\usepackage{enumitem}

\begin{document}
\title{Helping Code Reviewer Prioritize: Pinpointing Personal Data and its Processing\thanks{This paper has been accepted at the 22nd International Conference on Intelligent Software Methodologies, Tools and Techniques (SOMET2023).}}
\author{Feiyang Tang\inst{1} \and
Bjarte M. \O stvold\inst{1} \and
Magiel Bruntink\inst{2}}
\authorrunning{F. Tang et al.}
\institute{Norwegian Computing Center, Oslo, Norway \and
Software Improvement Group, Amsterdam, The Netherlands}
\maketitle              % typeset the header of the contribution
\begin{abstract}
Ensuring compliance with the General Data Protection Regulation (GDPR) is a crucial aspect of software development. This task, due to its time-consuming nature and requirement for specialized knowledge, is often deferred or delegated to specialized code reviewers. These reviewers, particularly when external to the development organization, may lack detailed knowledge of the software under review, necessitating the prioritization of their resources.

To address this, we have designed two specialized views of a codebase to help code reviewers in prioritizing their work related to personal data: one view displays the types of personal data representation, while the other provides an abstract depiction of personal data processing, complemented by an optional detailed exploration of specific code snippets. Leveraging static analysis, our method identifies personal data-related code segments, thereby expediting the review process. Our approach, evaluated on four open-source GitHub applications, demonstrated a precision rate of 0.87 in identifying personal data flows. Additionally, we fact-checked the privacy statements of 15 Android applications. This solution, designed to augment the efficiency of GDPR-related privacy analysis tasks such as the Record of Processing Activities (ROPA), aims to conserve resources, thereby saving time and enhancing productivity for code reviewers.

\keywords{Personal data processing \and GDPR analysis \and Static analysis \and Code review.}
\end{abstract}
\section{Introduction}
In the 21st century, companies have been collecting and distributing massive amounts of personal data~\cite{alharthi2017addressing}. The sheer volume and capacity to access and integrate data in unprecedented ways is novel~\cite{solove2001access}. To protect individual rights, the European Union's General Data Protection Regulation (GDPR) requires substantial data protection, posing both challenges and opportunities for global software companies and imposing severe fines for non-compliance~\cite{huth2019using}.
Article 30 of the GDPR necessitates the creation of a detailed document, the Records of Processing Activities (ROPA), to ensure GDPR compliance. This document, which must be readily accessible to the supervisory authority, can be challenging to construct. Code reviewers often play a crucial role in this process, tasked with analyzing extensive codebases for GDPR-relevant aspects. Their findings not only help ensure compliance but also assist in the formation of the intricate ROPA, a task that can be quite demanding.

While the manual identification of personal data and its processing in the codebase is an inevitable part of a code reviewer's work, our goal is to facilitate the rapid pinpointing and understanding of relevant code fragments. To this end, we propose an approach that identifies and abstracts both personal data and its processing in the codebase. Our solution provides two specialized views: one that emphasizes the types of personal data, and another that outlines an abstract perspective of data processing flows.

These views aim to empower code reviewers by streamlining the task of locating and comprehending personal data processing within the software. By providing a focused direction, highlighting key code fragments, and supplying task-specific information through different views, our approach conserves time and effort. This multi-faceted assistance greatly simplifies GDPR-related tasks, such as the formation of a ROPA.

The three main components of our approach are:
\begin{itemize}
    \item A set of adaptable static analysis rules for pinpointing personal data and its processing in source code. (Refer to Section~\ref{Sec:patternmatching}, \textit{identification} in Figure~\ref{fig:flowchart})
    \item A collection of flow patterns derived from large-scale analysis, which simplifies identified code fragments containing a data flow into abstracted code snippets. These snippets capture the context and manner in which personal data is processed. (Refer to Section~\ref{Sec:semantics}, \textit{abstraction} in Figure~\ref{fig:flowchart})
    \item Two specialized views provide code reviewers with options for displaying information about personal data or flow-specific information, depending on their specific task, thereby reducing manual work. (Refer to Section~\ref{Sec:SpecializedViews}, \textit{presentation} in Figure~\ref{fig:flowchart})
\end{itemize}

We demonstrate the effectiveness of our approach by 1) achieving high precision and generating corresponding ROPAs compared to published privacy statements for four trending GitHub projects and 2) evaluating the accuracy of privacy statements provided by 15 popular Android applications from the Google Play store.

\section{Challenges in Personal Data Identification}\label{Sec:pd}
Identifying personal data within extensive codebases poses a significant challenge for code reviewers, especially given the diversity of personal data types and the lack of standardized patterns for their identification. This issue is further complicated by the reality that personal data may be either directly or indirectly associated with an individual, a relationship that can vary across cultural, linguistic, and domain-specific contexts. Privacy statements often describe personal data collection and processing in broad terms, making it difficult to pinpoint what data is collected and how it is processed within the source code.

When reviewing code for GDPR-relevant aspects, code reviewers often resort to manual techniques such as keyword searching, filtering, and grepping to identify potential areas of concern. However, these methods can be time-consuming and often yield an overwhelming number of results, making it challenging to discern the key areas of focus. There is a clear need for a more abstract, categorized view of the results that could help reviewers identify and focus on the most relevant code fragments.

\paragraph{User Requirement Study}\label{Sec:userinterview}
In order to grasp the challenges faced by code reviewers in GDPR-related tasks, we undertook a user requirement study with six experienced code reviewers from a medium-sized European software company. They were selected based on their experience, familiarity with GDPR tasks, and diverse European representation.
The research, primarily focused on European data formats, did not consider formats outside the EU. The interviews elicited key issues in privacy analysis tasks, preferred presentation formats, and the potential of an abstract view of results.

During the structured interviews, participants discussed the most challenging facets of their privacy analysis tasks, the desired format for the presentation of findings, and their views on the possible advantages of having an abstract, categorized overview of the results.
The study highlighted that identifying personal data within source code remains a significant hurdle for reviewers. This challenge arises primarily due to the fluctuating context and semantics. Participants voiced a preference for a results presentation that offers ample context and underscores potential areas of concern, such as personal data processing.

There was unanimous agreement among participants about the potential benefits of incorporating an automated, abstract presentation of the results into their workflows. They preferred comprehensive coverage, even if it occasionally introduced false positives, on the condition that their manual examination could be concentrated on a smaller, more relevant segment of the software. While this study has certain limitations, such as a smaller sample size and potential bias due to the limited participant selection, it does provide valuable initial insights into the needs and preferences of code reviewers tasked with GDPR-related duties in Europe.

\paragraph{Presentation Design}
Current static analysis tools offer some flexibility in results grouping and presentation. For instance, tools like FindBugs and ESLint allow results to be grouped by various criteria, such as bug category/name, code location, and bug ranking. However, these tools often lack the ability to provide a task-specific view that aligns with the code reviewers' current analysis focus. Such a view could allow reviewers to focus on personal data and its processing, thereby simplifying the analysis process and making it more efficient.

\subsection{The Role of Code Reviewers in ROPA Creation}\label{Sec:preliminarymapping}

In the context of GDPR compliance, code reviewers often play a crucial role in creating a Record of Processing Activities (ROPA). A ROPA should contain detailed information about personal data processing, such as the categories of individuals and personal data, recipients of personal data, details of transfers to third countries, retention schedules, and technical and organizational security measures. Code reviewers, through their thorough analysis of the codebase, contribute significantly to the collection of this information, thereby helping to ensure GDPR compliance.

\section{Constructing Task-Specific Views for Code Reviewers}
We propose an approach that creates two distinct views, providing code reviewers with varying levels of information about personal data types and their processing within the codebase. Our approach employs static analysis to identify personal data and the source code where the processing of such data occurs. The result of the static analysis comprises identified code fragments that match the specified patterns. We introduce an abstracted flow representation that simplifies each identified code fragment, capturing the essence of personal data processing. This flow representation not only translates effectively into a more comprehensible format but also facilitates labeling and grouping of code fragments, presenting the results to code reviewers in a unified and efficient manner for analysis.

\begin{figure}[h!]
    \centering
    \begin{tikzpicture}[
        node distance=10mm and 25mm,
    box/.style = {draw, minimum height=15mm, text width=25mm, align=center,double,rounded corners},
    sy+/.style = {yshift= 2mm}, 
    sy-/.style = {yshift=-2mm},
    every edge quotes/.style = {align=center}
                            ]
    \node (n1) [box]             {Source code};                  
    \node (n2) [box,right=of n1] {Code fragments \\with flows};
    \node (n3) [box,below=of n2] {Simplified Code snippets};      
    \node (n4) [box,below=of n1] {Two views: \\personal data \\ and processing};
    \draw[thick,-Triangle]  
        (n1.east) to [above,"identification"] (n2.west);
    \draw[thick,-Triangle] 
        (n1.east) to [below,"(Section~\ref{Sec:patternmatching})"] (n2.west);
    \draw[thick,-Triangle] %dashed]  
        (n2.south) to [left,"abstraction\\(Section~\ref{Sec:semantics})"] (n3.north);
    \draw[thick,-Triangle]  
        (n3.west) to [above,"presentation"] (n4.east);
    \draw[thick,-Triangle]  
        (n3.west) to [below,"(Section~\ref{Sec:SpecializedViews})"] (n4.east);
    \end{tikzpicture}
    \caption{Overview of our approach}
    \label{fig:flowchart}
\end{figure}
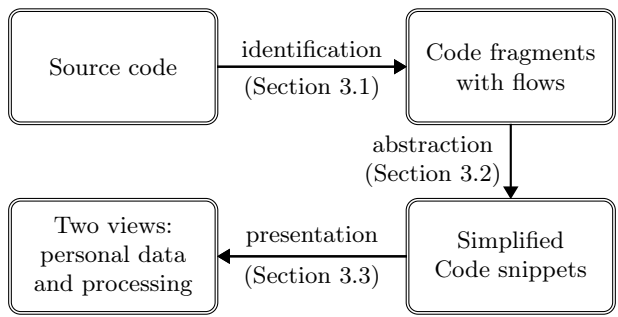

Figure~\ref{fig:flowchart} presents an overview of our approach, which consists of three major phases: pattern matching with static analysis, abstraction of code fragments to generate flow representations, and creation of the two task-specific views for code reviewers.

\subsection{Pattern Matching on Source Code}\label{Sec:patternmatching}

The first phase of our approach involves static analysis to pinpoint code fragments that contain or process personal data. For this task, we employ Semgrep, a powerful static analysis tool well-regarded for its flexibility and efficiency in analyzing voluminous source code files~\cite{naik2021sporq}. Semgrep's multi-language support and local data flow analysis capabilities are instrumental to our endeavor. This section elaborates on our rules for identifying personal data sources and sinks, and the subsequent extraction of flows.

\subsubsection{Identification of Personal Data and Processing}

The concept of sources and sinks is vital to our approach. In the context of our analysis, sources refer to personal data, while sinks represent different forms of personal data processing. We define personal data as 1) literal personal data in source code text (constants identified via real values), and 2) variables (identified via name identifier). As for personal data processing, it refers to any distinct action or operation performed on personal data.
Our rules for identifying personal data and its processing currently support Java, JavaScript, and TypeScript. Nevertheless, our rules applicable to plain-text personal data can be extended to the majority of Semgrep-supported languages. Semgrep parses the source code to build an abstract syntax tree (AST) for taint analysis, similar to how ESLint processes JavaScript code. This method enables us to efficiently identify sources, sinks, and data types. We further augment Semgrep's identification process with pattern matching.

\subsubsection{Crafting Identification Rules}

Literal personal data identification relies on matching specified regular expressions, such as the syntax of national ID numbers. For variable sources, we have established a default list of personal data identifiers, covering data from 10 categories: \textit{Account}, \textit{Contact}, \textit{Personal ID}, \textit{Online identifier}, \textit{Location}, \textit{Feedback}, \textit{Health}, \textit{National ID}, \textit{Technical}, and \textit{Financial}.

These distinct identifiers are used to construct Semgrep rules with regular expressions (regex). To reduce false positives and enhance recall, we apply restrictions to the regex rules. For instance, to identify all human names in source code, we improve precision and cover first, last, and full names by using regex such as \texttt{(?i).(?:first|given|full|last|sur(?!geon)\\)[s/(;)|,=!>]name)}.

Simultaneously, we identify potential sinks, which represent distinct actions of personal data processing. We utilize the majority of verbs from Section 3 of the Data Privacy Vocabulary (DPV)~\cite{pandit2019creating} to identify relevant processes. These verbs are utilized to generate corresponding regex for our taint analysis rules, and specific conditions are incorporated to pinpoint relevant sinks in the code effectively.

APIs are extensively used to implement functionalities in current software development. We performed a simple static analysis on the 20 most popular libraries each from Maven and npm (40 in total), followed by a manual inspection to filter out false positives. The resulting list consists of potential sink methods, predominantly related to databases from prominent providers such as AWS and Google Firebase. These API methods, akin to the verb identifiers, serve as sinks for our analysis. We categorize sinks into six classes: \textit{Manipulation}, \textit{Transportation}, \textit{Creation/Deletion}, \textit{Database}, \textit{Encryption}, and \textit{Log}.

\begin{table}
\centering
\caption{Source and sink category abbreviations}
\label{tab:abbr}
    \begin{tabular}{lrcclr}
    \toprule
    \textbf{Source type} & \textbf{Abbreviation} & & &\textbf{Sink type} & \textbf{Abbreviation} \\ \midrule
    Account              & ACC            & & &Manipulation       & M              \\
    Contact              & CON            & & &Transportation      & T              \\
    Personal ID          & PID            & & &Creation/Deletion  & C/D            \\
    Online identifier    & OID            & & &Database           & D              \\
    Location             & LOC            & & &Encryption         & E              \\
    Feedback             & FEE            & & &Log                & L              \\
    Health               & HEA            & & &                   &                \\
    National ID          & NID            & & &                   &                \\
    Technical            & TEC            & & &                   &                \\
    Financial            & FIN            & & &                   &                \\
    \bottomrule
    \end{tabular}
\end{table}

\subsubsection{Flow Extraction and Semgrep Output}\label{Sec:semgrepoutput}

Once sources and sinks have been identified, we need to understand how personal data flows from the former to the latter. A flow represents a sequence of operations, starting from a source, moving through intermediate nodes (if any), and ending at a sink. We extend Semgrep's capabilities to not only identify but also comprehend these flows.

Semgrep's output typically includes the code fragment where the flow ends, i.e., the statement that processes personal data. However, we have customized Semgrep to provide more detailed information, such as the precise location and name of the source and sink, along with the code fragment. This information is vital for code reviewers, helping them quickly locate and comprehend personal data processing within the software.

\paragraph{Example: Application on ToolJet}

To illustrate the effectiveness of our approach, we applied our rules to ToolJet, a low-code, open-source framework for creating and deploying internal tools. ToolJet was selected because of its popularity and extensive use of personal data, making it a representative case study. The ease of access to its source code also facilitated our analysis.

\begin{figure}[h!]
     \centering
     \begin{subfigure}{.45\textwidth}
         \centering
         \includegraphics[width=\textwidth]{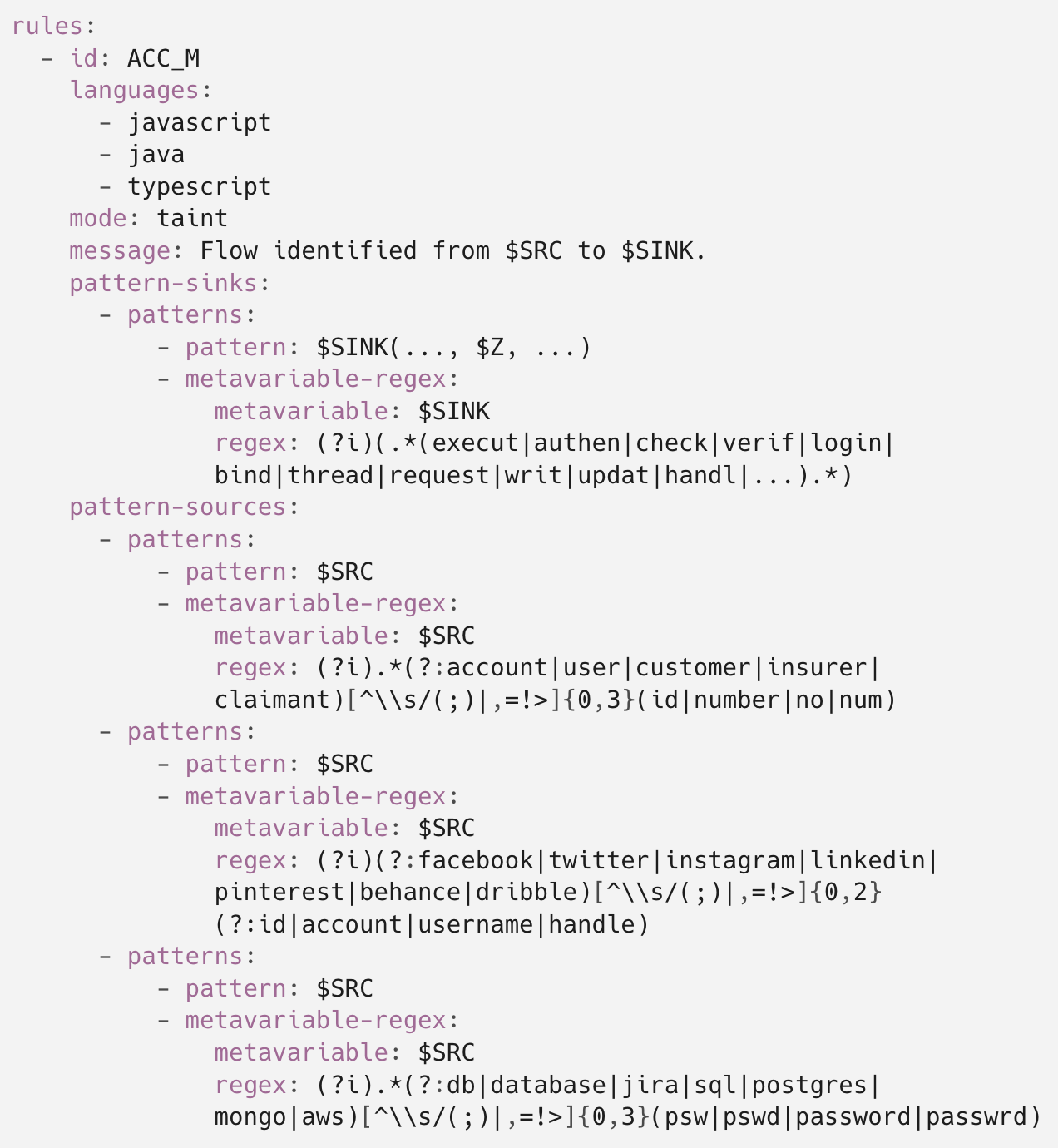}
         \caption{Sample identification rule to find flows between \textit{account} data and \textit{manipulation} sink}
         \label{fig:sample_rule}
     \end{subfigure}%
     \hspace{1em}
     \begin{subfigure}{.45\textwidth}
         \centering
         \includegraphics[width=\textwidth]{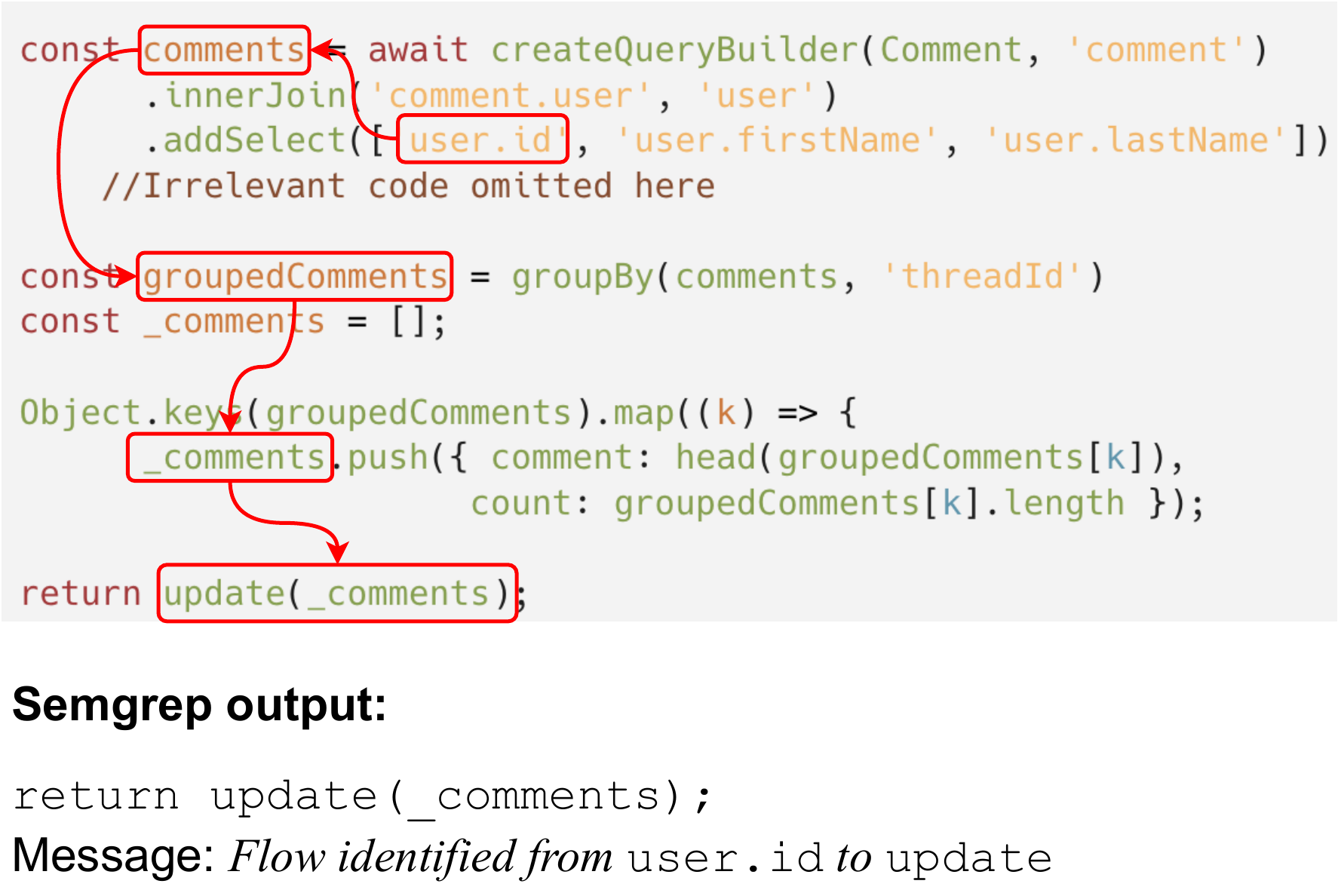}
         \caption{Code fragment identified using the rule illustrated in Figure~\ref{fig:sample_rule}. In this case, the flow originates from the source \texttt{user.id} and ends at the sink \texttt{update()}. Along with the code fragment, Semgrep provides an accompanying message that details the names of both the source and the sink involved in the flow.}
         \label{fig:identified_code_snippet}
     \end{subfigure}
        \caption{An example of pattern matching on ToolJet}
        \label{fig:patternmatch}
\end{figure}

Figure~\ref{fig:sample_rule} depicts an example of a rule we created to identify flows of \textit{account}-specific data into a \textit{manipulation} sink in ToolJet. The corresponding identified code snippet, an output of Semgrep that includes both the code fragment and key information about the source and sink, is shown in Figure~\ref{fig:identified_code_snippet}.

\subsection{Flow Patterns: Abstracting Personal Data Flows}\label{Sec:semantics}
A critical aspect of our approach is the abstraction of identified personal data flows into more manageable and comprehensible representations. This step facilitates the review and understanding of these flows, offering a more streamlined approach to software privacy analysis. Here, we introduce the concept of flow patterns, which are simplified, standardized representations of personal data flows, distilled from the identified code fragments.

Semgrep outputs a statement of code, which corresponds to a chunk of a code fragment that could span multiple lines in the source code, where the personal data flow culminates. The value at the sink within this statement may not necessarily be the original source, yet it definitely encapsulates the value originating from the source. However, such code fragments can be intricate and challenging to comprehend. In our work, we address this by abstracting each result we capture, which represents a flow from a personal data source to a processing sink. Our abstraction combines this code fragment with source and sink information to depict the flow in a simplified form.

We applied our pattern-matching rules to over 20 open-source software projects on GitHub (the top 20 trending ones across three languages: JavaScript, TypeScript, and Java) and identified a vast number of personal data flow instances. From these instances, we generalized the structure of more than 150,000 detected code snippets into a set of eight distinct flow patterns.

\subsubsection{Defining Flow Patterns}

The flow patterns, collectively denoted as $\mathcal{F}$, abstractly represent various forms of personal data flows from a source to or via a sink. Each pattern in $\mathcal{F}$ reflects a typical flow type, and Table~\ref{tab:flow_syntax} presents these flow patterns alongside their English interpretations. The table also introduces syntactic conventions for meta-variables used in the following discussion. $E$ ranges over expressions, $m$ ranges over methods, and $v$ ranges over source variables. An expression with an underscore as an argument, $E[\_]$, signifies that the expression is not significant in terms of personal data processing --- it is neither a source nor a sink, nor does it contain a value from a source.

In a flow pattern $E_1 \xrightarrow{m} E_2$, a solid arrow $\rightarrow$ indicates that a value on the left-hand side (LHS) contributes to the value on the right-hand side (RHS). If we are uncertain whether a value flows from the LHS to the RHS, such as when the LHS value is processed and merely outputs a Boolean to the RHS, we use a dashed arrow $\dashrightarrow$ to indicate that the LHS value may not fully reach the RHS. Each pattern's interpretation in English is also provided in Table~\ref{tab:flow_syntax}.

The flow patterns in $\mathcal{F}$ abstract the flows between sources and sinks occurring in the identified code snippets. Table~\ref{tab:example} offers eight examples illustrating the code snippets corresponding to flow patterns, ordered according to the flow patterns in Table~\ref{tab:flow_syntax}.

These examples represent the different types of flows that may occur when processing personal data in code snippets. By identifying, analyzing, and categorizing these flows into simplified flow patterns, we can provide a uniform view of the data flows in code, capturing the properties of sources and sinks in a single flow, the pathway from source to sink, and any other involved values. This abstraction step is crucial in creating an intuitive and efficient approach to understanding personal data flows, thereby enhancing privacy protection in software systems.

\begin{table}[t]
\caption{The collection of flow patterns $\mathcal{F}$}
\label{tab:flow_syntax}
\centering
    \begin{adjustbox}{width=\textwidth}
    \begin{tabular}{ll}
    \toprule
    \multicolumn{1}{l}{\textbf{Flow Patterns}}     & \multicolumn{1}{l}{\textbf{Translation to English}}                                \\ \midrule
    $E[\_] \xrightarrow{m} v $           & A non-personal data value flows via sink $m$ into a source $v$.                       \\
    $v_2 + E[\_] \stackrel{m}{\dashrightarrow} v_1$ & Value from source $v_2$ and non-personal data are both processed by sink $m$ and the result flows into source $v_1$. \\
    $v_2(E[\_]) \xrightarrow{m} v_1$                & Value from source $v_2$ flows via sink $m$ into source $v_1$.                                          \\
    $v \stackrel{m}{\dashrightarrow} E[\_]$          & Value from source $v$ gets processed by sink $m$ and the result flows into a new expression.                  \\
    $v \xrightarrow{m} E[\_]$                        & Value from source $v$ flows via sink $m$ into a new expression, making it a new source.                                                          \\
    $v_1 + v_2 \stackrel{m}{\rightarrow} v_1$    & Value from source $v_2$ flows via sink $m$ declared by $v_1$ into source $v_1$.                               \\
    $v + E[\_] \stackrel{m}{\rightarrow} v$         & Value from source $v$ and non-personal data are both processed by sink $m$ and the result flows into itself.                                               \\
    $v \stackrel{m}{\rightarrow} m(v)$           & Value from source $v$ flows into sink $m$.                                  
    \\ \bottomrule       
    \end{tabular}
    \end{adjustbox}
\end{table}

\begin{table}[t]
\caption{Examples demonstrating the relationship between code snippets and their corresponding flow pattern instances in $\mathcal{F}$. The examples are related to the flow patterns listed in Table~\ref{tab:flow_syntax}.}
\label{tab:example}
\centering
    \begin{adjustbox}{width=.7\textwidth}
    \begin{tabular}{ll}
    \toprule
    \multicolumn{1}{l}{\textbf{\textbf{Code Snippet}}} & \multicolumn{1}{l}{\textbf{\textbf{Flow Pattern Instance}}}                 \\ \midrule
    \texttt{full\_name = retrieve(record\_data,2)}     & E[\_] $\stackrel{\text{retrieve}}{\rightarrow}$ full\_name         \\
    \texttt{isFemale = check(user\_detail,'F')}        & user\_detail+ E[\_] $\stackrel{\text{check}}{\dashrightarrow}$ isFemale     \\
    \texttt{first\_name = UserInfo.get(2)}             & UserInfo(E[\_]) $\stackrel{\text{get}}{\rightarrow}$ first\_name            \\
    \texttt{choice = match(name,list)}                 & name $\stackrel{\text{match}}{\dashrightarrow}$ choice                 \\
    \texttt{choice = UserInfo.retrieve(2)}             & UserInfo $\stackrel{\text{retrieve}}{\rightarrow}$ choice                \\
    \texttt{AccountInfo.update(userId,index)}          & AccountInfo + userId $\stackrel{\text{update}}{\rightarrow}$ AccountInfo \\
    \texttt{AccountInfo.update(index)}                 & AccountInfo + E[\_] $\stackrel{\text{update}}{\rightarrow}$ AccountInfo     \\
    \texttt{print(SSN)}                                & SSN $\stackrel{\text{print}}{\rightarrow}$ print(SSN)              \\ \bottomrule     
    \end{tabular}
    \end{adjustbox}
\end{table}

\subsection{Specialized Views for GDPR Analysis Tasks}\label{Sec:SpecializedViews}

As GDPR compliance tasks can vary in their requirements, we provide specialized views to assist code reviewers in understanding and analyzing personal data handling within the system. An overview of personal data types and their distribution is useful when developing a broad understanding of personal data usage, while a detailed analysis of each data flow is required for tasks like constructing a Record of Processing Activities (ROPA) or assessing the risk of personal data handling. By providing these specialized views, we aim to support reviewers in efficiently and effectively performing GDPR compliance tasks.

\subsubsection{Personal Data Type View}\label{Sec:DataTypeView}

\begin{figure}[h!]
    \centering
    \includegraphics[width=0.5\textwidth]{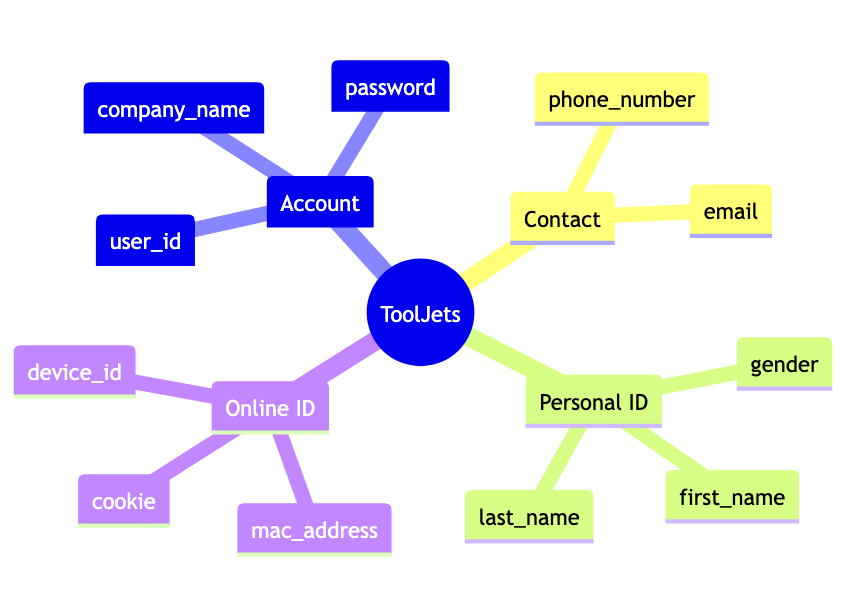}
    \caption{Personal Data Type View of ToolJet}
    \label{fig:pddist}
\end{figure}

The Personal Data Type View presents an overview of personal data identified in the source code. This view is a hierarchical representation that illustrates the distribution of personal data types. The tree-like structure is automatically generated based on our static analysis findings, categorizing personal data identifiers by their nature such as \textit{account} information or \textit{personal ID}. For instance, all source names with a stem-identifier like \texttt{email\_addr}, \texttt{email\_id}, \texttt{e-mail}, and \texttt{email} are grouped together. This overview provides users with a clear image of the types of personal data and their respective identifiers present in the source code. See Figure~\ref{fig:pddist} for an example.

\subsubsection{Detailed Flow View}\label{Sec:DetailedFlowView}

For tasks requiring a deeper understanding of specific personal data flows, we provide the Detailed Flow View. This view presents comprehensive information about each identified flow, including the file path, source and sink names and types, and abstract flow patterns. Furthermore, it offers the ability to link to the actual location of the code fragments, supporting a deeper contextual analysis when necessary. Features for filtering and ranking the identified flows are also included, allowing reviewers to focus on specific flows or those of high potential risk. An example of the Detailed Flow View is provided in Table~\ref{tab:emailmapping}.

These specialized views are designed to be flexible and adaptable to various GDPR compliance tasks. We incorporate grouping functionalities based on feedback from potential users, allowing results to be organized by specific source identifiers, file locations, or other relevant criteria. This flexibility is designed to make our approach practical, user-friendly, and adaptable to various GDPR compliance tasks.

\begin{table}[h!]
    \centering
    \caption{Detailed Flow View grouped by identifier \texttt{`email'} in ToolJet (top 7 results displayed)}
    \label{tab:emailmapping}
    \begin{adjustbox}{width=\textwidth}
        \begin{tabular}{lllcl}
        \toprule
        \multicolumn{1}{l}{\textbf{Path}}                 & \multicolumn{1}{l}{\textbf{Source}} & \multicolumn{1}{l}{\textbf{Sink}}     & \multicolumn{1}{l}{\textbf{Sink Type}} & \multicolumn{1}{l}{\textbf{Flow Pattern Instance}}                                       \\ \midrule
        server/src/services/organizations.service.ts      & users.email\_addr                   & createQueryBuilder                    & DB                                     & users.email\_addr $\stackrel{\text{createQueryBuilder}}{\rightarrow}$ query  \\
        server/src/services/group\_permissions.service.ts & users.email                         & createQueryBuilder                    & DB                                     & users.email $\stackrel{\text{createQueryBuilder}}{\rightarrow}$ query        \\
        server/src/services/users.service.ts              & email                               & this.usersRepository.findOne          & DB                                     & email+\_ $\stackrel{\text{findOne}}{\rightarrow}$ UserInfo                   \\
        server/src/services/organizations.service.ts      & email\_addr                         & this.usersService.create              & C/D                                    & email\_addr+\_ $\stackrel{\text{create}}{\rightarrow}$ UserInfo              \\
        server/ee/services/oauth/oauth.service.ts         & email                               & this.usersService.findOrCreateByEmail & C/D                                    & UserInfo+email $\stackrel{\text{findOrCreateByEmail}}{\rightarrow}$ UserInfo \\
        server/src/services/users.service.ts              & email                               & user.organizationUsers.sendData       & T                                      & email $\stackrel{\text{sendData}}{\rightarrow}$ sendData(email)              \\
        server/src/services/organizations.service.ts      & email\_addr                         & this.usersService.update              & M                                      & UserInfo+email\_addr $\stackrel{\text{update}}{\rightarrow}$ UserInfo   \\ \bottomrule
        \end{tabular}
    \end{adjustbox}
\end{table}

In a Detailed Flow View, we present key information such as file path, source and sink names and types, and flow pattern instances, which are displayed by default. For example, Table~\ref{tab:emailmapping} shows the Detailed Flow View for all source identifier \texttt{email} identification results. This information can be used to construct a ROPA using the official template provided by data protection authorities or a research semantic model like CSM-ROPA~\cite{ryan2021common}. By traversing flow pattern instances in Table~\ref{tab:emailmapping}, users can generate a list of processing related to personal data ``\texttt{email}'' and identify their location.

To better assist GDPR compliance, our Detailed Flow View should not be limited to a fixed presentation style. Based on feedback and recommendations from potential users, we incorporate grouping to simplify GDPR compliance. We also refer to the ICO's ROPA template~\cite{Howdowed33:online}, which specifies the information required for GDPR compliance mentioned in Section~\ref{Sec:preliminarymapping}, such as the categories of individuals and personal data, categories of recipients of personal data, transfer details to third countries, retention schedules, and technical and organizational security measures. Thus, we group the results by various attributes such as source and sink categories and their distinct identities, which can help users answer ROPA queries.

Moreover, users can experiment with different grouping criteria while examining the code. For example, they can group the results by a specific source identifier or file name/location of interest. We aim to incorporate GDPR-related grouping criteria to ease GDPR compliance and make our approach more useful.

\subsection{Implementation}\label{Sec:implementation}

In recognizing the contextual nature of personal data, we have designed a flexible system allowing users to customize pattern-matching rules via a Python script. This script simplifies the definition of flow syntax and the addition of new identifiers. All our identification rules are open-source to encourage collaborative enhancement of the system's capabilities.
The results of our analysis are produced in the Standard Static Analysis Results Interchange Format (SARIF), facilitating easy filtering and sorting based on parameters such as rules or file locations.

For visualizing the findings, we employ the Mermaid diagramming tool to automatically generate a personal data distribution graph. This visual representation aids in the identification and selection of specific source identifiers for further examination.
Finally, we've adopted a SARIF viewer for filtering and sorting functionalities, and designed an interactive interface for the Specialized Views, offering users an adjustable and task-specific data exploration experience.

\section{Experiment}
To assess our approach, we focus on open-source software that processes personal data. The experiment consists of two parts. In the first part, we analyze popular open-source projects from GitHub, manually evaluating the true positives (TP) and false positives (FP) in the results to validate our approach. We illustrate the usefulness of our two specialized views by aligning them with the published privacy statements of four validation applications, showcasing how these views can guide the creation of a ROPA. In the second part, we apply our approach to 15 popular Android applications on Google Play, scrutinizing the accuracy of the information given in the ``Data collected'' section of Google Play's ``Data safety'' section.

\subsection{Validation: Analyzing Trending GitHub Applications}
Since we have used ToolJet in the previous sections as an example, here we select our analysis target from the \textit{trending repositories on GitHub} list.
Among the repositories captured from the GitHub monthly trending list (accessed on 6/Dec/2022) under three different languages: Java\footnote{\url{https://github.com/trending/java?since=monthly}}, JavaScript\footnote{\url{https://github.com/trending/javascript?since=monthly}}, and TypeScript\footnote{\url{https://github.com/trending/typescript?since=monthly}}, we selected the top four complete applications, that are not a framework, an add-on, or a tutorial: Rocket Chat\footnote{\url{https://github.com/RocketChat/Rocket.Chat}}, Telegram\footnote{\url{https://github.com/DrKLO/Telegram}}, Odoo\footnote{\url{https://github.com/odoo/odoo}}, and Joplin\footnote{\url{https://github.com/laurent22/joplin}}.
We now refer to them as validation applications.

We display the time taken to complete the identification task on validation applications and the number of identified flows in Table~\ref{tab:cost}.

\begin{table}[htp]
    \caption{Identified code snippet count and the time consumed for each application}
    \label{tab:cost}
    \centering
    \begin{tabular}{lrr}
    \textbf{}            & \textbf{Time} & \textbf{No.\ of Code Snippet} \\ \hline\hline
    \textbf{Rocket Chat (TypeScript)} & 397\,s          & 6,935                 \\
    \textbf{Telegram (Java)}    & 562\,s          & 16,963              \\
    \textbf{Odoo (JavaScript)}        & 916\,s          & 25,653               \\
    \textbf{Joplin (TypeScript/JS)}      & 694\,s          & 17,299          \\\hline\hline  
    \end{tabular}
\end{table}

\begin{figure}[t]
\setkeys{Gin}{width=\linewidth}
    \begin{subfigure}{0.24\textwidth}
        \includegraphics[width=\textwidth]{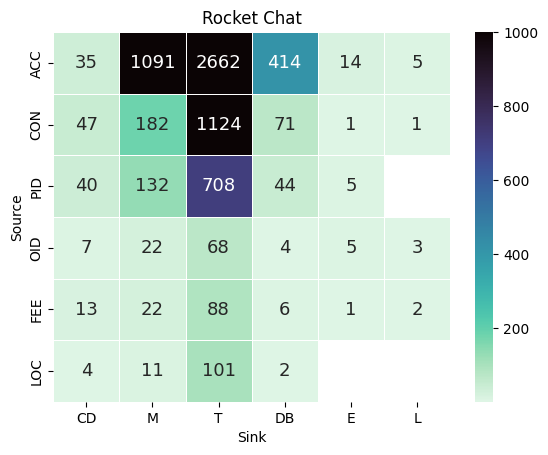}
        \caption{Rocket Chat}
        \label{fig:heatmap_rocketchat}
    \end{subfigure}
    \hfill
    \begin{subfigure}{0.24\textwidth}
        \includegraphics[width=\textwidth]{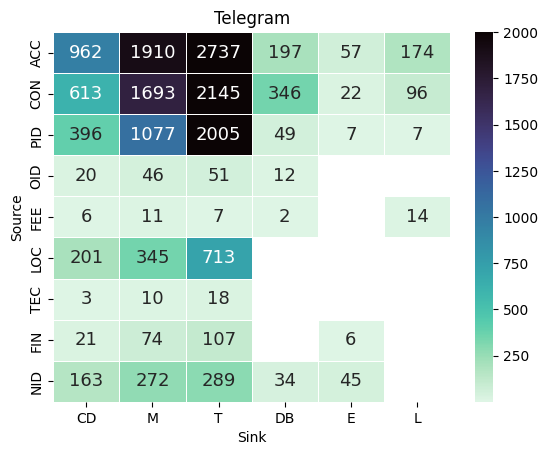}
        \caption{Telegram}
        \label{fig:heatmap_telegram}
    \end{subfigure}
    \medskip
    \begin{subfigure}{0.24\textwidth}
        \includegraphics[width=\textwidth]{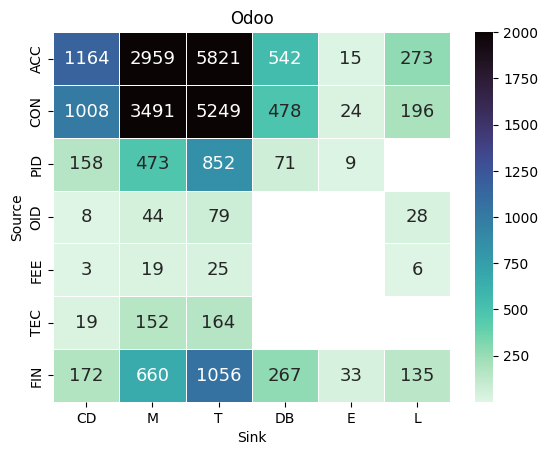}
        \caption{Odoo}
        \label{fig:heatmap_odoo}
    \end{subfigure}
    \hfill
    \begin{subfigure}{0.24\textwidth}
        \includegraphics[width=\textwidth]{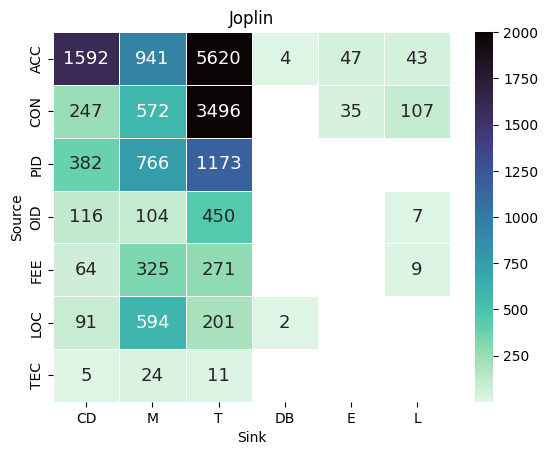}
        \caption{Joplin}
        \label{fig:heatmap_joplin}
    \end{subfigure}
    \caption{The overview statistics of identified flow under all possible combinations of source and sink types as illustrated in heatmaps. }
    \label{fig:heatmap}
\end{figure}

For large projects such as Odoo and Telegram, our analysis takes about 15 minutes to analyze 2,285 files (Odoo) using 70 static analysis rules.
Odoo is the business content management system with the highest number of identified personal data flows among the four validation applications, indicating that it processes a substantial amount of personal data.

Next, we validate our approach in two steps.
First, we manually examine the precision of identified code snippets, assessing whether they are related to personal data processing. Figure~\ref{fig:heatmap} and a Personal Data Type View figure (akin to Figure~\ref{fig:pddist}) form the initial perspective of our generated views. The four heatmaps present overview statistics of the different categories of personal data and its processing in the four validation applications. The overview statistics indicate that all four applications collect a significant quantity of account and contact information, which is understandable as they function as identifiers for system users and communication mediums. However, we observe that certain applications gather more categories of personal data from users, such as Telegram, which collects nine categories of personal data, with the majority stored in databases.

Table~\ref{tab:stats} summarizes the results of the code snippet detection in each example application, as well as the precision calculated through manual inspection for each identified flow provided in the Detailed Flow View. Given that this evaluation requires manual inspection and we need to avoid false negatives, we examine the code snippets identified by the static analysis using the following criteria to determine their precision score\footnote{Precision = TP/(TP+FP), where True Positives (TP) are identified results that meet both of our criteria, while False Positives (FP) are identified results that do not meet one of our criteria.}:

\begin{itemize}
\item Do the source and sink identifiers match their respective categories? For instance, if the analysis intended to match ``\texttt{log}'' but instead matched ``\texttt{login}'', which is irrelevant to the context.
\item Do the source and sink matches qualify as personal data processing? For instance, ``\texttt{noreply@test.org}'' was detected as a clear-text result for explicit personal data in the system, although it is not personal data.
\end{itemize}

\begin{table}[h!]
    \caption{Statistics showing the precision of different types of flows detected. `-' marks the labels for which our approach detected less than 20 results. Sink types are: \textit{data creation/deletion} (\textbf{C/D}), \textit{data manipulation} (\textbf{M}), \textit{data transportation} (\textbf{T}), \textit{database} (\textbf{DB}), \textit{encryption} (\textbf{E}) and \textit{log} (\textbf{L}). Source types are: \textit{account} (\textbf{ACC}), \textit{contact} (\textbf{CON}), \textit{personal ID} (\textbf{PID}), \textit{online identifier} (\textbf{OID}), \textit{location} (\textbf{LOC}), \textit{feedback} (\textbf{FEE}), \textit{health} (\textbf{HEA}), \textit{national ID} (\textbf{NID}), \textit{technical} (\textbf{TEC}), and \textit{financial} (\textbf{FIN}).}
    \label{tab:stats}
    \centering
    %\begin{adjustbox}{width=.5\textwidth}
    \begin{tabular}{lccccccc}
    \toprule
    \textbf{Application} & \textbf{Source} & \multicolumn{3}{c}{\textbf{Basic Sink}} & \multicolumn{3}{c}{\textbf{Special Sink}} \\
    \textbf{}            & \textbf{}       & \textbf{C/D}  & \textbf{M} & \textbf{T} & \textbf{DB}   & \textbf{E}  & \textbf{L}  \\ \midrule
    \textbf{Rocket Chat}& ACC & 0.80  & 0.95 & 0.94 & 0.97 & -    & -    \\
                      & CON & 0.85 & 0.88 & 0.95 & 0.96 & -    & -    \\
                      & PID & 0.83 & 0.92 & 0.95 & 0.96 & -    & -    \\
                      & OID & -    & 0.73 & 0.88 & -    & -    & -    \\
                      & FEE & -    & 0.82 & 0.91 & -    & -    & -    \\
                      & LOC & -    & -    & 0.89 & -    & -    & -    \\\hline
    \textbf{Joplin}   & ACC & 0.88 & 0.93 & 0.96 & -    & 0.89 & 0.95 \\
                      & CON & 0.93 & 0.81 & 0.92 & -    & 0.92 & 0.98 \\
                      & PID & 0.81 & 0.89 & 0.95 & -    & -    & -    \\
                      & OID & 0.79 & 0.87 & 0.84 & -    & -    & -    \\
                      & FEE & 0.84 & 0.95 & 0.8  & -    & -    & -    \\
                      & LOC & 0.91 & 0.86 & 0.89 & -    & -    & -    \\
                      & TEC & -    & 0.71 & -    & -    & -    & -    \\\hline
    \textbf{Telegram} & ACC & 0.89 & 0.92 & 0.96 & 0.87 & 0.86 & 0.96 \\
                      & CON & 0.90 & 0.96 & 0.94 & 0.93 & 0.91 & 0.92 \\
                      & PID & 0.84 & 0.93 & 0.89 & 0.86 & -    & -    \\
                      & OID & 0.70 & 0.73 & 0.81 & -    & -    & -    \\
                      & FEE & -    & -    & -    & -    & -    & -    \\
                      & LOC & 0.92 & 0.90 & 0.95 & -    & -    & -    \\
                      & TEC & -    & -    & -    & -    & -    & -    \\
                      & FIN & -    & 0.94 & 0.97 & -    & -    & -    \\
                      & NID & 0.86 & 0.96 & 0.94 & 0.88 & 0.91 & -    \\\hline
    \textbf{Odoo}     & ACC & 0.91 & 0.86 & 0.98 & 0.94 & -    & 0.95 \\
                      & CON & 0.84 & 0.87 & 0.93 & 0.92 & 0.92 & 0.96 \\
                      & PID & 0.82 & 0.95 & 0.97 & 0.96 & -    & -    \\
                      & OID & -    & 0.66 & 0.63 & -    & -    & 0.82 \\
                      & FEE & -    & -    & 0.72 & -    & -    & -    \\
                      & TEC & -    & 0.89 & 0.95 & -    & -    & -    \\
                      & FIN & 0.88 & 0.92 & 0.94 & 0.95 & 0.90  & 0.95 \\ \bottomrule
    \end{tabular}
    %\end{adjustbox}
\end{table}

Our main objective is to lessen the manual workload for code reviewers in identifying and analyzing personal data flows, though we recognize that manual scrutiny continues to play a crucial role. A high level of precision in our flow identification technique indicates that we can conserve resources by directing code reviewers to potential areas of concern, and offering guidance on the potential data flow paths and locations within their applications. This aligns with our specialized views approach, enhancing the efficiency and efficacy of GDPR compliance tasks.

For the manual analysis, we conducted a thorough evaluation of a representative sample of the identified code snippets to estimate the precision, rather than individually checking all code snippets identified.
Table~\ref{tab:stats} shows that the majority of flow types have a precision of at least 0.8, with an average of over 0.9 for categories with more than 500 identified flows. Some categories of flows that have a smaller sample size have a lower precision, ranging from 0.65 to 0.75. Although a precision of 0.8 might result in 5K false positives for an application with 25K identified snippets (e.g., Odoo), our approach aims to assist developers and code reviewers in finding possible directions and locations for personal data processing, thereby reducing the overall time and effort spent on manual analysis.

Note that we do not include the number of occurrences of literal personal data identified by the pattern matching, as their precision is typically less than 0.6.
A precision of less than 0.6 indicates that developers and code reviewers still need to devote considerable effort to exclude literal identification results that are not relevant.
Identifying literal personal data is difficult due to its ambiguous and highly contextual nature.

\begin{figure}[h!]
    \centering
    \includegraphics[width=.5\textwidth]{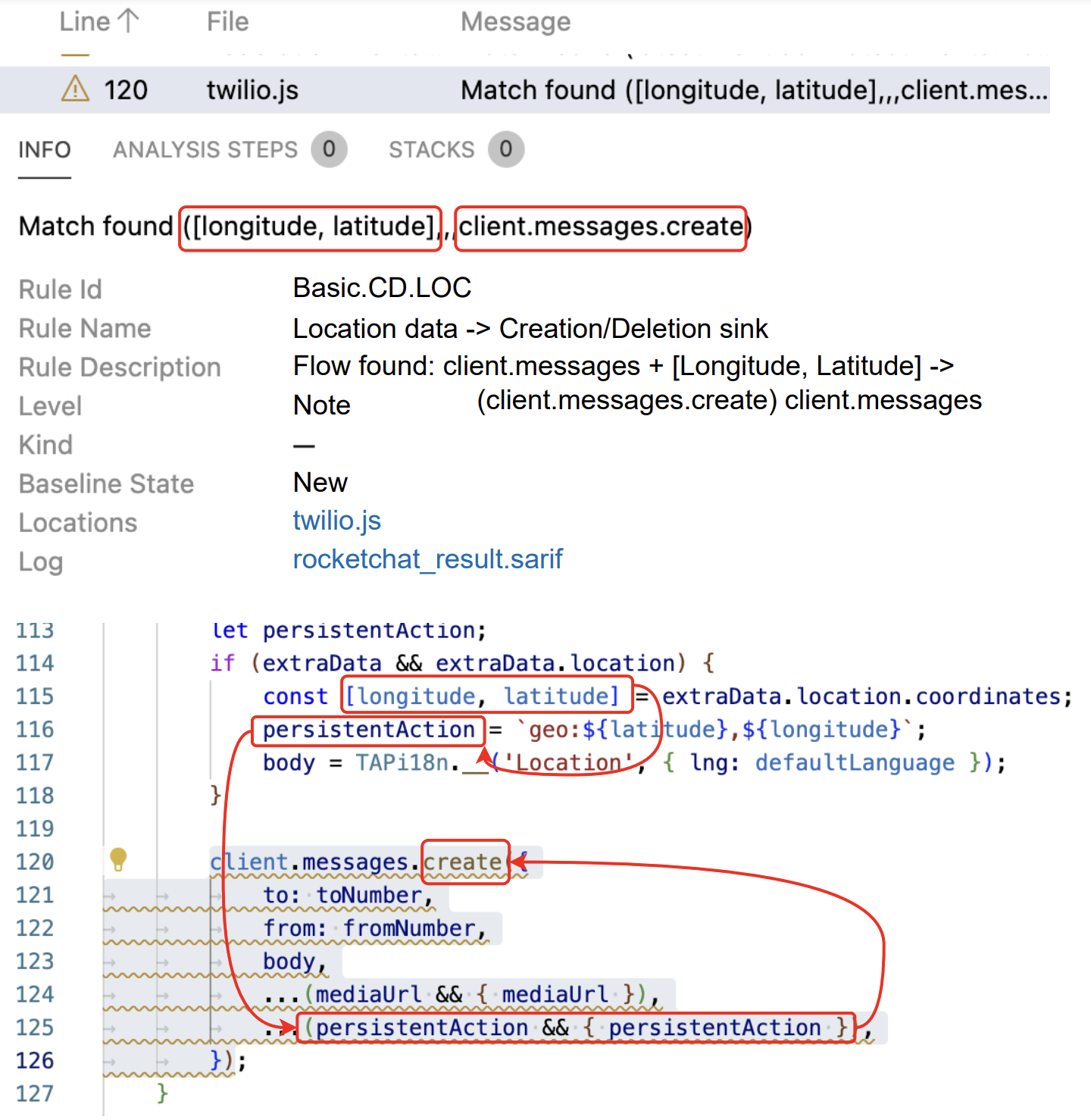}
    \caption{Example of an identified flow in the Detailed Flow View for Rocket Chat}
    \label{fig:map_rocketchat}
\end{figure}

Next, we leverage the attributes available in the detailed Detailed Flow View to evaluate existing privacy statements. Figure~\ref{fig:map_rocketchat} depicts an instance of a flow identified through our Detailed Flow View (similar to Table~\ref{tab:emailmapping}). This flow was identified by a rule pinpointing the flow of location data into a transportation sink. When a flow is selected for deeper scrutiny, crucial information such as the file location, rule type and name, abstract flow pattern (displayed under ``Rule Description''), source/sink identifiers, and the original code are readily accessible.

By integrating the Detailed Flow View with the provided statistics and heatmaps from the Personal Data Type View, we can guide ROPA development and verify the accuracy of existing privacy statements. For mobile applications, we additionally examine the accuracy of personal data processing disclosures made to platforms like Google Play or Apple's App Store. In situations where a comprehensive privacy statement is absent, such as with Joplin\footnote{\url{https://joplinapp.org/privacy/}}, we illustrate how our views can be employed for ROPA creation.

In order to establish a connection between ROPA requirements and our specialized views, we reference Table~\ref{tab:ropa} which addresses four critical requirements outlined in Section~\ref{Sec:preliminarymapping}. These can subsequently be incorporated into a comprehensive ROPA. Furthermore, we juxtapose our results with the published privacy statements of the chosen four applications.

Rocket Chat's detailed and well-structured privacy statement\footnote{\url{https://docs.rocket.chat/legal/privacy}} aligns closely with our Personal Data Type View and Detailed Flow View. Telegram and Odoo also offer comprehensive and lucid privacy policies. However, our experiment uncovered that Telegram collects user feedback and financial data, an activity not explicitly mentioned in their privacy statement.
On the other hand, Joplin's privacy statement\footnote{\url{https://joplinapp.org/privacy/}} is brief and lacks specificity on the types of personal data collected, except for geolocation data. We also discovered instances where some personal data is temporarily stored without being disclosed in the statement.

Our approach demonstrated its effectiveness in covering personal data and its processing in the selected validation applications, as compared to their published privacy statements. Nonetheless, we also found instances where the applications collected personal data beyond their disclosed policies. Personal data logging is a sensitive matter and should be meticulously inspected and documented.

\begin{minipage}{\textwidth}
\begin{center}
\captionsetup{type=table}
\captionof{table}{Fact-check the published privacy statements of the four applications using the information supplied by our two specialized views. The missing information is \underline{\textbf{highlighted}}.}
\label{tab:ropa}
\begin{adjustbox}{width=\textwidth}
\scriptsize
\begin{tabular}{p{0.15\textwidth}p{0.4\textwidth}p{0.45\textwidth}} \toprule
\textbf{Application} & \textbf{Data Captured via Views Aligned with ROPA} & \textbf{Published Privacy Statement/Policy} \\ \midrule
\textbf{Rocket Chat}
&
    \begin{itemize}
        \item Categories of personal data: ACC, CON, PID, OID, FEE, LOC.
        \item Categories of processing: basic processing: C/D, M, and T. Minor logging personal data identified.
        \item Transfer to a database or third-party APIs: own database access identified on ACC, CON, and PID data, minor on OID, FEE, and LOC data, no third-party database API detected.
        \item Data encryption or anonymization: Encryption on ACC, CON, PID, OID, FEE data. No anonymization was identified.
    \end{itemize}
&
    \begin{itemize}
        \item Categories of personal data: personal ID (PID, CON), account data (ACC), usage data (OID, FEE), location data (LOC), cookie data (OID)
        \item Categories of processing: for contact/identification: PID, CON; for market/communication: PID, CON; for registration: ACC; for maintenance/tech support/monitoring: OID, FEE; for functionalities: OID, LOC, OID.
        \item Transfer to a database or third-party APIs: no third-party services mentioned, data outside of the USA might be transferred to services in the USA.
        \item Data encryption or anonymization: relevant security measures were taken into account.
    \end{itemize}\\ %\hline
%%%
\textbf{Telegram}
&

\begin{itemize}
  \item Categories of personal data: ACC, CON, PID, OID, \underline{\textbf{FEE}}, LOC, TEC, \underline{\textbf{FIN}}, NID
  \item Categories of processing: basic processing: C/D, M, and T. Major logging ACC and CON data identified.
  \item Transfer to a database or third-party APIs: there are database calls for ACC, CON, PID, and NID data identified to both internal and external databases.
  \item Data encryption or anonymization: there is major encryption on NID, ACC, and CON data.
\end{itemize}
  &

\begin{itemize}
  \item Categories of personal data: account data (ACC, PID), contact data (CON, ACC, NID), location data (LOC), chats (OID), and cookies (TEC, OID)
  \item Categories of processing: for identification/account purposes (ACC, PID, NID); for communication (CON, ACC); for improving services (TEC, OID); for functionalities (LOC, TEC, OID).
  \item Transfer to a database or third-party APIs: data is saved in third-party provided data centers in the Netherlands for European users, and end-to-end chats are not transmitted out of the device
  \item Data encryption or anonymization: ``All data is stored heavily encrypted so that local Telegram engineers or physical intruders cannot get access.''
\end{itemize}\\ %\hline

%%%
\textbf{Odoo}
&

\begin{itemize}
  \item Categories of personal data: ACC, CON, PID, OID, FEE, TEC, FIN
  \item Categories of processing: basic processing: C/D, M, and T. Major logging ACC, CON, and FIN data identified.
  \item Transfer to a database or third-party APIs: there are database calls for ACC, CON, PID, and FIN data identified to both internal and external databases.
  \item Data encryption or anonymization: there is encryption on ACC, CON, PID, and FIN data.
\end{itemize}
  &
\begin{itemize}
  \item Categories of personal data: account \& contact Data (ACC, CON), job application data (CON, PID), browser data (FEE, TEC), customer databases (ACC, CON, PID, FIN), free trial session recording (FIN, TEC, OID, FEE), In-App Purchase transaction data (FIN)
  \item Categories of processing: for the recruitment process (ACC, CON, PID); for maintaining and improving services (FEE, TEC, OID); for providing services (ACC, PID), answering requests (CON), and for billing management (FIN, CON, ACC)
  \item Transfer to a database or third-party APIs: ``customer databases are hosted in the Odoo Cloud Region closest to where they are based, and can request a change of region''
  \item Data encryption or anonymization: ``info is securely processed, stored and preserved from data loss and unauthorized access''.
\end{itemize}\\ %\hline

%%%

\textbf{Joplin}
&
\begin{itemize}
  \item Categories of personal data: \underline{\textbf{ACC}}, \underline{\textbf{CON}}, \underline{\textbf{PID}}, \underline{\textbf{OID}}, \underline{\textbf{FEE}}, LOC, \underline{\textbf{TEC}}
  \item Categories of processing: basic processing: C/D, M, and T. Major logging contact data identified.
  \item Transfer to a database or third-party APIs: Almost none, only less than 5 identified for ACC and LOC data to be passed into a local temporary data model.
  \item Data encryption or anonymization: There are encryption measures on ACC and CON data identified.
  \end{itemize}
  &
\begin{itemize}
  \item Categories of personal data: It is not mentioned in the privacy policy, only explicitly mentioned geo-location data.
  \item Categories of processing: Not mentioned.
  \item Transfer to a database or third-party APIs: Only mentioned: ``Any data that Joplin saves, such as notes or images, are saved to your own device and you are free to delete this data at any time.''
  \item Data encryption or anonymization: Not mentioned.
\end{itemize}\\ \bottomrule

\end{tabular}
\end{adjustbox}
\end{center}
\end{minipage}

\subsection{Application: Assessing Google Play Data Safety Statements}

To further highlight the applicability of our approach, we examined the top three Android applications from Google Play (accessed on 8/Dec/2022, from the U.S. store\footnote{\url{https://play.google.com/store/apps?gl=US}}) in five different categories: communication, online shopping, weather, fitness/health, and transportation.
As the decompilation process generated some noise, potentially distorting the count of data processing flows, we focused on assessing the categories and distributions of personal data collected by the applications.

In our evaluation, we verified the ``Data Safety'' statements of 15 applications against the Personal Data Type View and Detailed Flow View generated by our approach. Table~\ref{tab:androidresults} presents the results, including the number of identified flows for each personal data category, the coverage of their data safety statements in Google Play, and the amount of personal data logged.

\begin{table}[tb]
\caption{Identified personal data types, their quantity, and how they are covered by Google Play's privacy policies. In column `Coverage', `-' indicates the types of personal data identified by our approach but not covered in the privacy statement; `+' implies that there is coverage of all identified personal data types in the privacy statement.}
\label{tab:androidresults}
\begin{adjustbox}{width=\textwidth}
\begin{tabular}{lrrrrrrrrrrrr}
\toprule
\multicolumn{1}{l}{\textbf{Android Apps}} &
  \textbf{ACC} &
  \textbf{CON} &
  \textbf{PID} &
  \textbf{OID} &
  \textbf{LOC} &
  \textbf{FEE} &
  \textbf{HEA} &
  \textbf{NID} &
  \textbf{TEC} &
  \textbf{FIN} &
  \textbf{Coverage} &
  \textbf{Logging} \\ \midrule
\textbf{Discord}             & 1621 & 791  & 31   & 399  & 159  & 107 & -    & -  & 87  & 9   & +                 & 41  \\
\textbf{Kik}                 & 2065 & 1923 & 17   & 817  & 51   & 39  & -    & -  & 46  & 31  & -ACC, -LOC           & 133 \\
\textbf{Viber}               & 1823 & 2093 & 101  & 740  & 219  & 42  & -    & -  & 71  & 16  & -ACC, -OID, -FEE, -TEC & 82  \\
\textbf{Amazon}              & 905  & 2762 & 1025 & 1623 & 634  & 198 & 42   & -  & 231 & 129 & -ACC                & 162 \\
\textbf{AliExpress}          & 1244 & 2381 & 1730 & 2094 & 841  & 217 & -    & -  & 309 & 261 & -LOC, -FEE           & 217 \\
\textbf{Shein}               & 1072 & 1967 & 965  & 978  & 409  & 96  & -    & -  & 160 & 373 & -LOC                & 62  \\
\textbf{The Weather Channel} & 239  & 2384 & 1291 & 1328 & 2623 & 172 & -    & -  & 16  & 22  & +                 & 76  \\
\textbf{AccuWeather}         & 176  & 2137 & 1102 & 1736 & 3321 & 98  & -    & -  & 5   & 9   & -PID, -OID, -ACC, -CON & 41  \\
\textbf{Windy}               & 312  & 246  & 567  & 803  & 1196 & -   & -    & -  & -   & 14  & -LOC                & 22  \\
\textbf{SamSung Health}      & 273  & 182  & 31   & 98   & 685  & 44  & 1027 & -  & 28  & 21  & +                 & 10  \\
\textbf{Google Fit}          & 105  & 49   & 14   & 6    & 271  & 19  & 580  & -  & 9   & 4   & +                 & 7   \\
\textbf{My Fitness Pal}      & 328  & 209  & 19   & 113  & 902  & 75  & 721  & -  & 47  & 69  & -PID                & 89  \\
\textbf{Uber}                & 1497 & 1054 & 298  & 835  & 2094 & 511 & -    & -  & 814 & 427 & +                 & 294 \\
\textbf{Trainline}           & 469  & 107  & 244  & 54   & 626  &     & -    & 4  & 82  & 562 & -NID                & 104 \\
\textbf{Omio}                & 724  & 151  & 457  & 72   & 729  &     & -    & 12 & 205 & 710 & -ACC, -NID, -OID      & 71  \\ \bottomrule     
\end{tabular}
\end{adjustbox}
\end{table}

We found that our approach detected certain types of personal data not disclosed in the Google Play data safety section. While Google allows exceptions for ``on-device access/processing'' and ``end-to-end encryption'', we found instances where data left the device without being disclosed. For example, location data in the online chatting app Kik and national ID data in the transportation app Omio are transmitted outside the device, but this was not stated in their respective Google Play disclosures.

Interestingly, we observed that applications across various domains commonly collected account and contact data, with additional domain-specific types. For instance, weather applications gathered significant location data, while transportation/ticketing applications acquired national ID data from users. Consistent with the GitHub projects evaluated earlier, all the apps logged personal data. This assessment underscores the utility of our approach in helping apps improve the accuracy of their data safety statements and increase transparency in their data handling practices.

\subsection{Threats to Validity}

Our approach does not offer a fully automated solution for GDPR compliance tasks, such as generating a ROPA, due to the lack of a natural language processing module. This limitation makes it challenging to directly generate or verify statements based on the output of our approach, which consists of code snippets, fragments, labels, and additional details. As a result, manual effort is required in our experiments, which consequently restricts the number of applications we can feasibly validate (four in this study). Each application must be open-source, widely used, gather diverse types of personal data, and possess a publicly available privacy statement.

The primary threat to the validity of our experiments is the difficulty in establishing a ground-truth set of actual data processing activities within the source code. This is due to the necessity of domain knowledge from the original development team, which is typically inaccessible. Consequently, it is not feasible to accurately measure the recall value (TP/(TP+FN)) for our approach.

In the context of fact-checking Google Play data safety statements, we are unable to verify which types of personal data are sent outside of the device. This limitation prevents us from fully assessing the accuracy of these statements.

\section{Related Work}
Existing research on incorporating privacy-by-design (PbD) into the software development life cycle lacks concrete tools to assist software developers in designing and implementing GDPR-compliant systems~\cite{baldassarre2020integrating}. Moreover, such PbD-created systems lack standards for mapping particular legislative data protection obligations, such as the GDPR.

Ferrara et al. propose a framework to adopt static analysis to assess GDPR compliance~\cite{ferrara2018static}. However, this approach requires compiled bytecode, which may not always be readily available during the development process. Arfelt et al. provide a formal logic for monitoring GDPR compliance~\cite{arfelt2019monitoring} and a functional tool was developed by~\cite{9328756} for analyzing cross-border data transmission in Android applications, which is limited in scope to a specific platform.

Some existing research focuses on the identification of personal data but does not consider broader GDPR compliance aspects. Fugkeaw et al.\cite{fugkeaw2021ap2i} design a technique to let enterprises automatically detect and handle personal data stored in the local file system. ReCon\cite{ren2016recon} uses machine learning to detect possible breaches of personal data by monitoring network traffic, requiring a pre-trained ML model. van der Plas~\cite{van2022detecting} identifies personal data in Git commits using CodeBERT, a transformer model similar to RoBERT, but this approach only focuses on the presence of personal data.

Our work aims to address these research gaps by providing a comprehensive solution that does not rely on compiled bytecode, pre-trained ML models, or platform-specific tools, and goes beyond merely identifying the presence of personal data to ensure broader GDPR compliance during the software development process.

\section{Conclusion and Future Directions}

Ensuring GDPR compliance demands detailed information on personal data processing, often requiring significant manual effort. Our work strives to lessen this burden by offering two specialized views --- Personal Data Type View and Detailed Flow View, facilitating code reviewers in identifying potential data processing locations and providing necessary information. Our approach, leveraging static analysis, has shown an average precision of 0.87 in our experiments, demonstrating its effectiveness.

However, our approach has limitations. Currently, it is based on Semgrep for static analysis, which captures intra-procedural data flows, leaving inter-procedural flows unaccounted for. The adoption of the Semgrep Pro Engine, offering inter-procedural analysis, could enhance our approach's precision. %Furthermore, our identification rules hinge on predefined regular expressions and identifiers, indicating scope for further refinement and potentially increased precision, possibly by analyzing data models in the source code.

In conclusion, our approach presents specialized views, aiding in GDPR compliance tasks, such as ROPA production, by reducing manual effort. With further improvements, such as incorporating the Semgrep Pro Engine and refining our identification rules, we aim to make the process of privacy analysis more efficient and manageable for code reviewers.

\section*{Acknowledgement}
This paper is an extended version of work published in~\cite{icissp23}. We would like to extend our sincere gratitude to Rob van der Veer for his valuable insights and contributions to this research. This work is part of the Privacy Matters (PriMa) project. The PriMa project has received funding from European Union’s Horizon 2020 research and innovation program under the Marie Skłodowska-Curie grant agreement No. 860315.

 \bibliographystyle{splncs04}
 \bibliography{ref}
\end{document}